\documentclass{appolb}
\usepackage{graphicx}

\begin{document}
\title{
Low-energy spin fluctuations of the heavy-Fermion compound 
CeNi$_2$Ge$_2$: origin of non-Fermi liquid behavior
}


\author{
H. Kadowaki$^{(1)}$, T. Fukuhara$^{(2)}$, K. Maezawa$^{(2)}$
\address{
$^{(1)}$ Department of Physics, Tokyo Metropolitan University, 
Hachioji-shi, Tokyo 192-0397, Japan\\
$^{(2)}$ Department of Liberal Arts and Science, 
Toyama Prefectural University, 
Kosugi, Toyama 939-0398, Japan
}
}
\maketitle


\begin{abstract}
Neutron scattering shows that non-Fermi-liquid behavior of the 
heavy-Fermion compound CeNi$_2$Ge$_2$ is brought about by 
development of low-energy spin fluctuations with an energy scale 
of 0.6 meV. 
They appear around antiferromagnetic wave vectors 
$(\frac{1}{2} \frac{1}{2} 0)$ and $(0 0 \frac{3}{4})$ 
at low temperatures, 
and coexist with high-energy spin fluctuations with an energy scale of 
4 meV and a modulation vector 
$(0.23, 0.23, \frac{1}{2})$. 
The energy dependence of the spin fluctuations is a peculiar 
character of CeNi$_2$Ge$_2$ which differs from typical heavy-Fermion 
compounds, 
and suggests importance of low-energy structures of quasiparticle bands. 
\end{abstract}


\vspace{0.4 cm}
Non-Fermi-liquid (NFL) behavior has been investigated in a number of 
$f$-electron systems in recent years 
\cite{Stewart01}. 
In heavy-Fermion systems, large mass enhancements are brought about 
by fluctuations of the spin degrees of freedom participating in 
quasiparticles. 
When certain spin fluctuations are slowed down, 
the Fermi liquid (FL) description breaks down and NFL behaviors 
appear. 
For many NFL compounds, the slowing down is related to proximity to 
magnetic phases or quantum critical points (QCP), 
$T_{\mathrm{N}} =0$, 
which requires tuning by chemical substitutions or static pressures. 

A typical NFL compound CeCu$_{5.9}$Au$_{0.1}$ has been intensively studied 
experimentally \cite{Schroder00}. 
The NFL behaviors of this system have also been discussed 
theoretically from viewpoints of QCP, where 
an important question is whether the critical singularity 
is explained by the standard spin fluctuation theory of an itinerant 
character \cite{Moriya} or by 
a fixed point with a localized spin character \cite{Coleman01}. 
Unfortunately, inevitable disorder effects on QCP 
by chemical tuning give some difficulty in the interpretation of NFL. 
Among many NFL compounds a few examples, 
such as CeNi$_2$Ge$_2$ \cite{Steglich96-Julian96} 
and YbRh$_2$Si$_2$ \cite{Trovarelli00}, 
are stoichiometric and provide opportunities of studying 
NFL behaviors or QCP in clean limits. 
The compound CeNi$_2$Ge$_2$ belongs to paramagnetic 
heavy-Fermion compounds with enhanced $C/T \simeq 350$ mJ/K$^2$mole 
and $T_{\mathrm{K}} \simeq 30$ K \cite{Knopp88}, 
implying similarity to the typical FL CeRu$_2$Si$_2$. 
However precise measurements at low temperatures, 
for example $C/T \propto \ln (T_0 / T)$, 
revealed that CeNi$_2$Ge$_2$ 
shows NFL behaviors with an energy scale much smaller than 
$T_{\mathrm{K}}$ \cite{Steglich96-Julian96}. 
In this study, we directly measured magnetic excitations of 
CeNi$_2$Ge$_2$ using neutron scattering and have found 
low-energy spin fluctuations which are the origin of the NFL behaviors. 

A single-crystalline sample of 2.2 cm$^3$ in volume was grown using the isotope 
${}^{58}$Ni to avoid the large incoherent elastic scattering 
of natural Ni nuclei, which is essentially important for 
observing spin excitations in a low energy range. 
Neutron scattering experiments were performed on the triple-axis 
spectrometer ISSP-HER installed at JRR-3M JAERI (Tokai). 
The typical energy resolution using a fixed 
$E_{\mathrm{f}}= 3.1$ meV condition was 
0.1 meV (FWHM) at the elastic position.

\begin{figure}[!ht]
\begin{center}
\includegraphics[width=0.9\textwidth,clip]{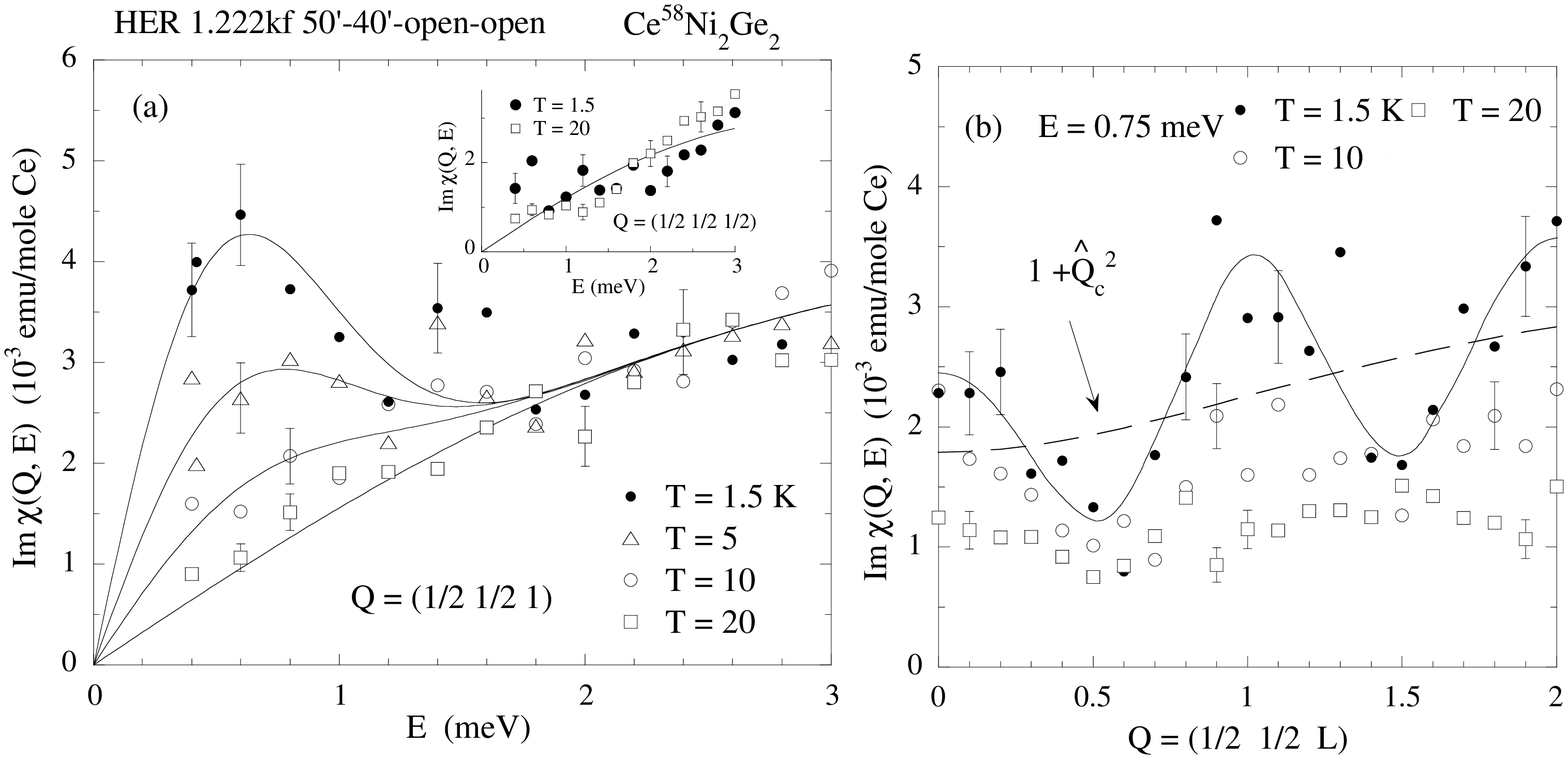}
\end{center}
\caption{
(a) Constant-$Q$ scans at 
$\mathbf{Q} = (\frac{1}{2} \frac{1}{2} 1)$. 
Solid lines are fits to the Lorentzian 
with $\Gamma_{\mathbf{Q}} = 4$ meV for data at $T = 20$ K 
and the function of Eq.~(\ref{eq:L+G}) with an additional 
Gaussian for data below 10 K. 
The inset shows constant-$Q$ scans at 
$\mathbf{Q} = (\frac{1}{2} \frac{1}{2} \frac{1}{2})$ 
and a fit curve to the Lorentzian. 
(b) Constant-$E$ scans along a line 
$\mathbf{Q} = (\frac{1}{2} \frac{1}{2} L)$ 
taken with $E = 0.75$ meV. 
Solid and dashed lines are a guide to the eye and an orientation factor 
of spin fluctuations in the $ab$-plane, respectively, 
for data at 1.5 K.
}
\label{fig:EQscan}
\end{figure}

It has been found that a pronounced spin fluctuation grows around a 
wave vector $\mathbf{Q} = (\frac{1}{2} \frac{1}{2} 1)$. 
Constant-$Q$ scans at this wave vector in the temperature range $1.5<T<20$ K 
are shown in Fig.~\ref{fig:EQscan}(a), together with those at 
$\mathbf{Q} = (\frac{1}{2} \frac{1}{2} \frac{1}{2})$. 
At 20 K the both data are well described by the Lorentzian form 
$\mathrm{Im} \chi_{\mathrm{L}} (\mathbf{Q}, E) = 
\chi_{\mathrm{L}}(\mathbf{Q}) E \Gamma_{\mathbf{Q}} 
/ ( E^2 + \Gamma_{\mathbf{Q}}^2 ) $ 
with $\Gamma_{\mathbf{Q}} \simeq 4$ meV $= 44$ K in agreement with the previous 
neutron scattering experiments \cite{Knopp88,Fak-JPCM}. 
The spin fluctuation of this energy scale is antiferromagnetic 
with a characteristic wave vector 
$\mathbf{k}_1 = (0.23, 0.23, \frac{1}{2})$ \cite{Fak-JPCM}. 
The same spectral shape persists down to 1.5 K for 
$\mathbf{Q} = (\frac{1}{2} \frac{1}{2} \frac{1}{2})$. 
On the other hand at 
$\mathbf{Q} = (\frac{1}{2} \frac{1}{2} 1)$, 
the spectral weight in a low energy range increases below 10 K. 
Since the NFL behaviors of CeNi$_2$Ge$_2$ show salient features in 
this $T$ range, we can conclude that this low-energy spin fluctuation 
is relevant to NFL. 
It should be noted that its characteristic wave vector 
$\mathbf{k}_2 = (\frac{1}{2} \frac{1}{2} 0)$ 
is near the antiferromagnetic wave vector of 
Ce(Ni$_{1-x}$Pd$_{x}$)$_2$Ge$_2$, 
indicating the proximity to an antiferromagnetic phase. 
Since the observed spectral shapes in Fig.~\ref{fig:EQscan}(a) show 
an additional peak structure below 1.5 meV, we tried to parameterize 
the data by adding either another Lorentzian or a Gaussian term 
to $\mathrm{Im} \chi_{\mathrm{L}} (\mathbf{Q}, E)$. 
Only the latter form 
\begin{equation}
\label{eq:L+G}
\mathrm{Im} \chi (\mathbf{Q}, E) = 
\mathrm{Im} \chi_{\mathrm{L}} (\mathbf{Q}, E) + 
\delta \chi (\mathbf{Q}) 
\frac{\sqrt{\pi} E }{ \gamma_{\mathbf{Q}} } 
\mathrm{e}^{ - (E/\gamma_{\mathbf{Q}})^2 } ,
\end{equation}
provides reasonable fits, which are shown by solid lines in 
Fig.~\ref{fig:EQscan}(a). 
The energy width of the Gaussian is 0.65 meV (HWHM) at $T$ = 1.5 K. 

In Fig.~\ref{fig:EQscan}(b) we show constant-$E$ scans taken with 
$E = 0.75$ meV along a line $\mathbf{Q} = (\frac{1}{2} \frac{1}{2} L)$ 
below and above 10 K. 
From this figure one can see that a periodic modulation with 
peaks at integral $L$, implying an antiferromagnetic correlation 
with the wave vector $\mathbf{k}_2$, is formed only at low temperatures. 
Second information extracted from these data is an anisotropy of 
the spin fluctuation. 
The slow $Q$ variation of the intensity is brought about by 
an orientation factor $(1 + \hat{Q}_{c}^{2})$ of the spin fluctuation 
in the $ab$-plane as shown by the dashed curve in Fig.~\ref{fig:EQscan}(b). 

\begin{figure}[!ht]
\begin{center}
\includegraphics[width=0.6\textwidth,clip]{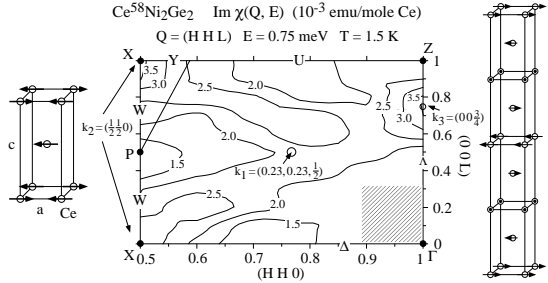}
\end{center}
\caption{
A contour map of constant-$E$ scans taken with $E = 0.75$ meV in the $(HHL)$ 
scattering plane at $T = 1.5$ K. 
In hatched area, data were not observed owing to non-magnetic noise. 
Antiferromagnetic spin configurations, depicted in left and right 
sides, illustrate low-energy spin fluctuations with wave vectors 
$\mathbf{k}_2 = (\frac{1}{2} \frac{1}{2} 0)$ ($X$ point) and 
$\mathbf{k}_3 = (0 0 \frac{3}{4})$, respectively, 
assuming spins along the $a$-axis. 
The wave vector $\mathbf{k}_1 = (0.23, 0.23, \frac{1}{2})$ is 
the position where the high-energy spin-fluctuation 
($\Gamma_{\mathbf{Q}} \simeq 4$ meV) shows the 
maximum intensity \cite{Fak-JPCM}.
}
\label{fig:map}
\end{figure}

A number of constant-$E$ scans were carried out at 1.5 K 
in a wider $Q$ range, to check the interesting possibility of 
two-dimensional (2D) spin fluctuations inferred from the resistivity exponent 
and to find out other spin fluctuations. 
A typical result is shown in Fig.~\ref{fig:map} as a contour 
map of intensities with $E = 0.75$ meV in the $(HHL)$ scattering plane. 
In this figure $\mathbf{k}_2$ corresponds the two $X$ points 
$(\frac{1}{2} \frac{1}{2} 0)$ and $(\frac{1}{2} \frac{1}{2} 1)$. 
A simple rod-type scattering along $(\frac{1}{2} \frac{1}{2} L)$ of 
the 2D fluctuation has not been observed. 
There is, instead of this, another peak structure at a reduced wave vector 
$\mathbf{k}_3 = (0 0 \frac{3}{4})$, which also grows only at low temperatures. 
Constant-$Q$ scans at $\mathbf{k}_3$ show similar spectra 
to those of $\mathbf{k}_2$ with a slightly larger energy width of the Gaussian term of 0.9 meV (HWHM), which suggests secondary importance of 
$\mathbf{k}_3$ to NFL. 

It has been elucidated that a basic characteristic of CeNi$_2$Ge$_2$ is 
the existence of the low and high energy scales of the spin fluctuations 
with different $Q$ dependences. 
In stark contrast with this, spin fluctuations of standard 
heavy-Fermion compounds, such as CeRu$_2$Si$_2$ and CeCu$_6$, 
possess continuous energy scale $\Gamma_{\mathbf{Q}}$ of the Lorentzian 
$\mathrm{Im} \chi_{\mathrm{L}}$ modulated by RKKY interactions. 
For NFL CeCu$_{5.9}$Au$_{0.1}$ chemical tuning reduces 
$\Gamma_{\mathbf{Q}}$ further to zero at the antiferromagnetic wave vector. 
On the other hand, the NFL behavior in CeNi$_2$Ge$_2$ is caused by 
the formation of the second low-energy spin-fluctuations, 
which is the Gaussian term of Eq.~(\ref{eq:L+G}). 
Although the reason why this second spin fluctuation appear 
is not clear, at present, we may speculate that certain fine structures of quasiparticle 
bands, ex. nesting, formed well below $T_{\mathrm{K}}$ lead to the 
low-energy structure of spin excitations.

\end{document}